Electronic spectroscopy of a cold SiO$^+$ sample: Implications for optical pumping

Patrick R. Stollenwerk and Brian C. Odom

Department of Physics and Astronomy, Northwestern University,

2145 Sheridan Road, Evanston IL 60208, USA

and

Damian L. Kokkin, and Timothy Steimle

School of Molecular Sciences, Arizona State University

Tempe, Arizona 85287-1604, U.S.A.

Number of pages: 18
Number of tables: 1
Number of figures: 4

**Abstract**

The dispersed fluorescence following pulsed dye laser excitation of the $B^2\Sigma^+$- $X^2\Sigma^+(0,0)$ band of a cold sample of $SiO^+$ has been recorded and analyzed. The branching ratios for $B^2\Sigma^+(v=0) \rightarrow X^2\Sigma^+(v)$ and $B^2\Sigma^+(v=0) \rightarrow A^2\Pi_i(v)$ emission were determined and compared with values predicted based upon existing experimental and theoretical data. The experimentally determined branching ratios show that the $B^2\Sigma^+(v=0) \rightarrow X^2\Sigma^+(v)$ transitions are somewhat less diagonal than predicted. The implications for laser cooling of a trapped sample of $SiO^+$ using broadband laser excitation are discussed.

## I. Introduction

Recently there has been an effort to extend laser techniques used for cooling internal and external degrees of freedom of atoms to molecules. This effort is motivated by the many advantages molecules have over their atomic counterparts for metrology, quantum information and simulation, and ultracold chemistry applications[1-3]. A prerequisite for many applications is the ability to spontaneously scatter a large number of photons over a short time period. Unlike atomic systems, molecules can be removed from the optical cycling scheme either by fluorescing to other vibrational levels or pre-dissociation due to spin-orbit and rotational mixing of excited electronic states. Although the lack of vibrational selection rules precludes cycling an electronic transition as perfectly as is common in laser-cooled atoms, a select set of molecules with near diagonal Franck-Condon Factors (FCFs) allow for many optical-frequency photons to be resonantly scattered before a vibrational state change breaks the cycle. Examples of molecules with near diagonal FCFs include the neutral species of SrF[4-6] and YO[7], whose properties have been exploited for direct laser cooling and magneto-optical trapping, and CaF, which has been directly laser cooled[8]. AlH[+], whose rotational population has been cooled with the use of lasers[9], is an example of a ionic molecule. Hamamda et al.[10] recently reviewed the progress and prospects of internal state cooling and control of diatomic neutral and ionic molecules.

Molecular trapping and optical cooling technologies are important for many applications, and unique advantages can be found on either side of the choice to work with either neutral or charged molecules. As ion traps offer deep trapping potentials, extremely long trapping lifetimes, and excellent control over systematic uncertainties, it is prudent to identify ionic species with optical properties favorable for laser cooling experiments. The deep trapping potentials and ability to co-load laser-cooled atomic ions that can sympathetically cool a room temperature molecular ion's translational degrees of freedom down to the motional ground state of the ion trap[11] relaxes the requirements of a favorable molecular structure. This is because only 10 to 100 photon scatters are necessary for internal cooling compared to the roughly $10^4$ photon scatters necessary for Doppler cooling. Using quantum logic spectroscopy[12], externally and internally cooled molecular ions are able to be probed non-destructively[13] in a field-free environment ideal for precision measurements.

Broadband lasers are of special interest for internal cooling of molecules because a single laser source contains the multiple spectral components needed to address the large number of initially populated states in a room temperature molecule. Spectral filtering of the broadband laser can be used to remove pieces of the spectrum which would lead to internal heating. This technique was first demonstrated by vibrationally cooling $Cs_2$[14]. It was also proposed that for a transition with diagonal FCFs, applying a sharp cutoff at the right frequency such that $P$-branch transitions are predominately driven will cause rotational cooling[15-17]. Rotational cooling by this method was first demonstrated[9] with trapped AlH+. Other candidates with favorable properties for possible laser cooling in an ion trap were identified by Nguyen et al[15, 18], amongst them includes the focus of this study, the silicon monoxide cation, SiO+.

The design of laser cooling experiments requires both an understanding of the molecular quantum levels of the ground and excited electronic states as well as excited state decay paths and probabilities. For the simplest implementation of broadband laser rotational cooling, near diagonal FCFs are important so that vibrational excitation is not likely to occur during the cooling process, although in some cases it should also be possible to re-pump vibrational excitations[15, 18]. In the case of SiO+ the relevant states are the $X^2\Sigma^+$, low-lying $A^2\Pi$ ($T_e$ =2242 cm$^{-1}$), and $B^2\Sigma^+$ ($T_e$ =26029 cm$^{-1}$) states. The quantum levels of the $X^2\Sigma^+$, $A^2\Pi$, and $B^2\Sigma^+$ states have been investigated primarily by Rosner's group using fast-ion beam methods[19-23]. In addition to electronic spectroscopy[19, 20, 22, 24], magnetic dipole allowed transitions between fine structure components of numerous rotational levels in the $X^2\Sigma^+$ (v=0) state were also measured by Rosner's group using a separated field optical pump/probes rf-double resonance technique[21]. In their work, SiO+ was generated by applying a discharge to heated silicon monoxide powder and subsequently accelerating the charged products at 8 kV into a beam that passes through a Wien filter that allows the products to be separated by mass. By overlapping the molecular beam with a tunable , frequency doubled, cw - Ti:sapphire laser, they were able to measure $B^2\Sigma^+$ state excited state lifetimes and assign 2378 lines of the laser induced fluorescence (LIF) spectrum for the $X^2\Sigma^+$(v=0-9)$\rightarrow$ $B^2\Sigma^+$(v=0-5) and $A^2\Pi$ (v=0-10)$\rightarrow$ $B^2\Sigma^+$(v=0-5) electronic transitions. Optical transitions associated with very high rotational levels (J > 80) were detected due to the elevated temperature of the fast ion-beam sample. Optical features associated with the lowest quantum levels, which are relevant to laser cooling experiments, were not assigned. The full data set was fitted to a model Hamiltonian, precisely determining the molecular constants for the $X^2\Sigma^+$, $A^2\Pi$, and $B^2\Sigma^+$ electronic states and the $X^2\Sigma^+/A^2\Pi$ interaction parameters[22]. The determined spectroscopic parameters were used by Reddy *et. al.* [25] to predict the FCFs in support of optical spectroscopic searches for $SiO^+$ in the atmospheres of cool S-type stars. The radiative branching ratios and transition moments have not been experimentally determined. In this paper we determine and/or place limits on the radiative branching ratios for the $B^2\Sigma^+$ (v=0) excited state following laser excitation of a selected branch feature in $B^2\Sigma^+$ - $X^2\Sigma^+$(0,0) band of a cold free-jet expansion sample of $SiO^+$. To the best of our knowledge, this is the first reported radiative branching ratio measurement of a cold molecular ion using LIF detection. A method for generating a cold molecular beam sample of $SiO^+$ is also described.

Due in a large part to its importance in energetic, Si-rich environments (e.g. plasma and circumstellar envelopes), there have been high-level theoretical predictions[26-30] of the ground and low-lying electronic states of $SiO^+$. Cai reported on the *ab initio* prediction of the potential energy curves (PECs) and spectroscopic parameters for the $X^2\Sigma^+$ and $A^2\Pi$ states[26], the PECs, spectroscopic parameters, for the $B^2\Sigma^+$ and the $B^2\Sigma^+$ - $X^2\Sigma^+$ transition moment as a function of internuclear separation [27]. Most recently this group predicted PECs, spectroscopic parameters, and electric dipole moments for eight bound doublet states as well as electric dipole allowed transition moments between these states[28]. An internally contracted multi-reference configuration interaction approach was used. Relativistic and spin-orbit interactions were not treated. A somewhat higher–level *ab initio* prediction for the $X^2\Sigma^+$, $A^2\Pi$, and $B^2\Sigma^+$ state was performed[29] soon thereafter using multi-reference singles and doubles configuration interaction method which included relativistic and spin-orbit coupling interactions. The most relevant facet of that theoretical study to the present report was the prediction of the $B^2\Sigma^+$ - $X^2\Sigma^+$ and $A^2\Pi$ - $X^2\Sigma^+$ transition moments as a function of internuclear separation. The most recent *ab initio* calculation[30] was designed to improve the quality of the PECs and spectroscopic parameters curves for the eight lowest electronic states (four doublet states and four quartet states) by inclusion of core-valence correlation and relativistic corrections. Transition probabilities were not predicted.

## II. Experimental

The experimental setup is nearly that used in our previous experimental detection and characterization of the SiHD radical[31] and that used for branching ratio determination of ThO[32]. Three types of measurements were performed: two-dimensional (2D) spectroscopy, dispersed fluorescence (DLIF) spectroscopy, and radiative decay. In the present study, a cold sample of $SiO^+$ was produced via a pulsed d.c.-discharge (40 µs, 1kV, 1kOhm) struck through a supersonic expansion reaction mixture of silane ($SiH_4$) (0.8%), nitrous oxide ($N_2O$) (0.05%) in argon (600psi). The resulting free-jet expansion containing $SiO^+$ was subsequently probed 10 cm downstream with the output of an excimer pumped dye laser over the range of 26000-26050 cm$^{-1}$ covering the $B^2\Sigma^+$- $X^2\Sigma^+$(0,0) band system. A commercial wavemeter was used to determine the absolute wavenumber of the pulsed dye laser. Laser induced fluorescence (LIF) was imaged via adaptive optics onto the entrance slit of a 0.67m fast (f=6.2) monochromator equipped with a 300 lines/mm grating. The monochromator was fitted with a gated, intensified CCD (ICCD) camera cooled to -30ºC to reduce dark current.

In order to identify and isolate the $SiO^+$ LIF signal from the multitude of other molecules generated in the d.c.-discharge source a two-dimensional (2D) spectroscopic technique[33, 34] was employed similar to that used previously to study SiHD[31] and ThO[32]. Briefly, the 2D spectra are created by stepping the probe laser wavelength and capturing a 75 nm wide spectral region of the dispersed laser induced fluorescence (DLIF) for each laser wavelength step. The entrance slit widths were set to 1 mm resulting in an approximately ±2 nm spectral resolution for the DLIF signal. In this way the laser excitation spectrum of $SiO^+$ could be identified due to its characteristic, well resolved, DLIF signal.

Following the initial 2D survey scans, higher resolution, more sensitive, DLIF measurements following $B^2\Sigma^+$- $X^2\Sigma^+$(0,0) excitation were taken by tuning the laser wavelength to be on resonance with a branch feature and accumulating a large number (15000) of ICCD exposures for a contiguous set of 75 nm spectral windows across the 380-470nm range. In this case the entrance slit of the monochromator was narrowed to 0.3 mm resulting in a spectral resolution of ± 0.3 nm. Wavelength calibration of the resulting DLIF spectrum was achieved by measuring the emission from an argon pen lamp. The conjoined spectra were flux calibrated for the detection system. Similarly, a background spectrum was recorded by tuning the probe laser off resonance. Both photon counting and the more traditional current mode processing of the ICCD signal were employed. For photon counting, a photon signal threshold was set such that background dark counts were minimal while still accumulating sufficient signal counts. As the fluorescence signal rate averaged much less than one detected photon per probe pulse, the photon counting mode was found to improve the signal-to-noise ratio (SNR) over conventional operation of the ICCD. The comparatively low signal rate is expected as coulomb repulsion will result in densities of $SiO^+$ that is several orders of magnitude lower than what can be found in typical neutral beam densities.

The $B^2\Sigma^+$ (v=0) excited state lifetime measurements were performed by sitting the pulsed dye laser on an $R(3)$ line of the $B^2\Sigma^+$- $X^2\Sigma^+$(0,0) band monitoring the DLIF spectrum with a relatively wide (1μs) intensifier detection window. The detection window was progressively stepped further in time from the incident pulsed laser in 2ns increments. The resulting fluorescence decay curve was then fit with a first order exponential to determine the upper state lifetime.

### III. Results and Analysis

Figure 1 shows the 2D spectrum of the $SiH_4/N_2O/Ar$ discharge covering the 26000 cm$^{-1}$ to 26035 cm$^{-1}$ region of the $B^2\Sigma^+$ - $X^2\Sigma^+$ (0,0) transition of $SiO^+$. The horizontal axis is the laser excitation frequency and the vertical axis is the dispersed fluorescence wavenumber relative to the wavenumber of the laser. The center of the approximately 75nm spectral window monitored by the monochromator was offset approximately 1700 cm$^{-1}$ to the red with respect to the laser energy and tracked with the laser. The 2D spectrum of Figure 1 has been cropped to display only the shortest wavelength 40 nm portion of the 75nm spectral window. A sum of 50 discharge pulses at each laser excitation frequency was accumulated. The regular pattern of features, indicated by the yellow rectangle, appearing on-resonance are primarily the *P*- and *R*-branch structure of the $B^2\Sigma^+$- $X^2\Sigma^+$ (0,0) band of $SiO^+$. The observed spectral resolution of approximately 0.4 cm$^{-1}$, which is dictated by the bandwidth of the laser, is insufficient to resolve the spectral doubling due to spin-rotation splitting in the $B^2\Sigma^+$ and $X^2\Sigma^+$ states. There is no obvious feature in the 2D spectrum associated with the $B^2\Sigma^+$ (v=0) → $X^2\Sigma^+$ (v=1) emission, which would appear at 1162 cm$^{-1}$ (=$\Delta G_{1/2}$) on the DLIF axis. This off-resonance emission is observed in the DLIF spectra (see below), which were recorded using a significantly larger number of pulses (50 vs. 15000). The two very intense, partially resolved, doublets that appear in the 2D spectrum near the laser excitation wavelengths of 26008 cm$^{-1}$ and 26027 cm$^{-1}$ are due to laser excitation of the ${}^{S}R_{21f}(4.5)(\nu=26008.3\,\text{cm}^{-1})$, ${}^{S}R_{21e}(4.5)(\nu=26008.7\,\text{cm}^{-1})$, ${}^{S}R_{21f}(5.5)(\nu=26026.5\,\text{cm}^{-1})$ and ${}^{S}R_{21e}(5.5)(\nu=26026.9\,\text{cm}^{-1})$ branch features of the A$^2\Delta$-X$^2\Pi_r$ (1,0) transitions of SiH[35]. The intense, off-resonance, DLIF signal in the 2D spectrum at approximately 2050 cm$^{-1}$ red shifted from the laser are primarily the emission associated with the intense ${}^{R}R_{1f}(4.5)(\nu=23963.1\,\text{cm}^{-1})$, ${}^{R}R_{1e}(4.5)(\nu=23963.5\,\text{cm}^{-1})$, ${}^{R}R_{1f}(5.5)(\nu=23969.5\,\text{cm}^{-1})$ and ${}^{R}R_{1e}(5.5)(\nu=23970.3\,\text{cm}^{-1})$ transitions of the A$^2\Delta$ -X$^2\Pi_r$ (1,1) band[35]. The carriers of the remaining numerous weak features in the 2D spectrum are unknown.

The one dimensional excitation spectrum obtained by vertical integration of the horizontal slice indicated by the yellow rectangle of Figure 1 is displayed in Figure 2 (upper trace). The lower trace in Figure 2 is the simulated laser excitation spectrum obtained using the previously determined constants for the B$^2\Sigma^+$(v=0) and X$^2\Sigma^+$(v=0) states[22], an effective rotational temperature of 40 K and a linewidth of 0.4 cm$^{-1}$ FWHM, which is commensurate with laser bandwidth. The line indicated by the arrow is due to an unidentified molecule that emits both on-resonance and approximately 490 cm$^{-1}$ off-resonance, as evident in the 2D spectrum. The $P$(4), and higher members of that branch, are partially overlapped and blended with weak emission from an unknown molecule which contributes to the slight disagreement between the observed and predicted relative intensities. With this in mind, it can be concluded that the current production scheme of pulsed d.c.-discharge supersonic expansion is producing a cold ($\cong$ 40K $\pm$ 5K) molecular sample of SiO$^+$.

DLIF spectra were recorded using both the conventional monitoring mode and a photon counting mode of the ICCD acquisition software. Figure 3 shows the normalized DLIF spectra recorded using the conventional method for 18 sets of 15000 discharge pulses and the photon counting method for 12 sets of 15000 pulses. In both cases the laser was tuned to be on resonance with the overlapped $R_{11}$(3) (v=26022.1691cm$^{-1}$)[19] and $R_{22}$(3) (v=26022.1289 cm$^{-1}$)[19] branch features, which is the most intense line. Despite the larger acquisition of the spectrum obtained using the convention method, the signal-to-noise (SNR) ratio of the spectrum recorded using the photon counting mode is significantly improved. The large gain of the ICCD allows for the photon count threshold to be well above the dark counts and read noise, removing both sources of noise, without sacrificing signal. Exclusion of dark counts also flattens the baseline as temperature gradients lead to uneven dark count rates across the pixel array. Dark count rates are also subject to drift with changing environmental conditions which can explain the variation of baselines seen in the conventional ICCD acquisition mode plot in Figure 3(b). DLIF spectra recorded when the laser was tuned off-resonance midway between the $P(1)$ and $R(0)$ lines and when the excitation laser was blocked were used to measure the background contributions.

The DLIF spectra were used to estimate the branching ratios, $b_{v',v''}$, which are reported in Table 1. Peaks observed at the expected transition locations were fit to a Gaussian line shape with a linear offset using the method of least squares. Multiplication of the amplitude and width of the optimized Gaussian line shape was used to assign relative signal strengths and error propagation of the covariance matrix of the fit parameters was used to estimate the statistical uncertainty. Observed background peaks from scattered laser and discharge emission were subtracted away from the signal. After extracting signal sizes, the signals are normalized with respect to the detector's wavelength response before calculating branching ratios. For expected transitions where no peaks were observed (e.g. $B^2\Sigma^+$ (v=0) $\rightarrow$ $A^2\Pi_{1/2}$ (v=1) (v~23780cm$^{-1}$ , $\lambda_{vac}$~ 420 nm)) the data was fit to a straight line and RMS of the linear fit used as an upper bound on peak heights. Multiplying the peak height limit by the width of observed peaks places a limit on signal size and consequently the branching ratio. In assigning error bars to the branching ratio, the upper-side uncertainty comes from the statistical uncertainty of the fit parameters. The lower-side uncertainty comes from the statistical uncertainty added in quadrature with the one-sigma limit on the unobserved fluorescence from $B^2\Sigma^+\rightarrow$ $A^2\Pi$ spontaneous emission. This uncertainty assumes that branching to higher vibrational levels of the $X^2\Sigma^+$ state are negligible and, as the calculations predict, that the $B^2\Sigma^+$ (v=0) $\rightarrow$ $A^2\Pi$(v=0,1,2) transitions are much more intense than other members of this series. No assumptions about the $B^2\Sigma^+\rightarrow$ $A^2\Pi$ electronic transition moment are made.

As indicated in Table 1, and seen in Figure 3, excitation of the $B^2\Sigma^+$- $X^2\Sigma^+$ (0,0) band resulted in detection of only the strong on-resonance fluorescence and a much weaker $B^2\Sigma^+$ (v=0)→ $X^2\Sigma^+$(v=1) emission. Fluorescence branching is observed only to $X^2\Sigma^+$ with all other channels remaining undetected. As the decrease in $b_{iv',fv''}$ from v''=0 to v''=1 is so large, the limit placed on v''=2 is likely much larger than the true value and the limit should also extend to the sum over all higher vibrational transitions. No branching to the $A^2\Pi$ state was observed, consistent with predictions of a smaller transition dipole moment[18, 36, 37] (see below).

The fluorescence lifetime of the $B^2\Sigma^+$ state of $SiO^+$ following excitation was measured. Figure 4 displays the integrated fluorescence decay curve for excitation of the overlapped $R_{11}(3)$ and $R_{22}(3)$ branch features. The resulting curve was then fit with a single exponential function giving a determined lifetime of $66 \pm 2$ ns. This compares well with the previously determined value of $69.5 \pm 0.6$ ns obtained using an alternative method[20], confirming that the carrier of the spectrum is $SiO^+$.

## IV. Discussion

The primary purpose of this branching ratio measurement is to supplement the currently available data of the energy level structure and excited state lifetimes in order to design a broadband laser cooling scheme applicable to trapped $SiO^+$. As mentioned, the well spectrally separated *P*, *R*-branch structure of the $B^2\Sigma^+$- $X^2\Sigma^+$ (0,0) band facilitates rotational cooling via broadband laser source modified with a simple spectral mask[15]. This rotational cooling scheme relies on repeated removal of rotational quanta via the *P*-branch of the $B^2\Sigma^+$-$X^2\Sigma^+$ (0,0) transition, while at the same time avoiding having molecules fluoresce to levels other than the $X^2\Sigma^+$(v=0). Indeed, the determined branching ratios, $b_{iv',fv''}$(exp.) demonstrate that $B^2\Sigma^+$ (v=0)→ $X^2\Sigma^+$(v) emission is near-diagonal and $SiO^+$ fulfills these criteria.

The observation that the $B^2\Sigma^+$ (v=0) → $X^2\Sigma^+$(v) emission is near diagonal and much stronger than the $B^2\Sigma^+$ (v=0) → $A^2\Pi$(v) emission is readily qualitatively explained using a simple molecular orbital correlation diagram or the results of the theoretical prediction[27, 29]. The primary configuration for the $X^2\Sigma^+$ and $A^2\Pi_i$ states are $1\sigma^2 2\sigma^2 3\sigma^1 1\pi^4$ and $1\sigma^2 2\sigma^2 3\sigma^2 1\pi^3$ respectively, whereas the $B^2\Sigma^+$ state is an admixture of the $1\sigma^2 2\sigma^2 3\sigma^0 1\pi^4 4\sigma^1$ (51%) and $1\sigma^2 2\sigma^2 3\sigma^0 1\pi^3 4\sigma^1 2\pi^1$(18%) configurations. Here only the 2*s*(O), 2*p*(O), 3*s*(Si) and 3*p*(Si) atomic orbitals are considered in labelling the molecular orbitals. To a first order approximation $2\sigma$, $1\pi$, $4\sigma$, and $2\pi$ are the $2p$(O)$\pm$ $3p$(Si) bonding and anti-bonding orbitals with the $2\sigma$ and $1\pi$ being strongly polarized towards O and the $4\sigma$ and $2\pi$ towards Si. The $1\sigma$ and $3\sigma$ are the non-bonding $2s$(O) and $3s$(Si) orbitals respectively. Simultaneous $3s$(Si) and $3p$(Si) participation in bonding is inhibited because of the large difference in radial extent. The $X^2\Sigma^+ \rightarrow B^2\Sigma^+$ transition is associated with a charge transfer, one electron promotion, whereas the $A^2\Pi_i \rightarrow B^2\Sigma^+$ transition is a much weaker, two electron, promotion. Hence, the $X^2\Sigma^+ \rightarrow B^2\Sigma^+$ electric transition dipole moment is much larger ($\sim$1.83 D at $R_e$) than that for the $A^2\Pi_i \rightarrow B^2\Sigma^+$ ($\sim$0.10 D at $R_e$) (see below). Furthermore, the $X^2\Sigma^+ \rightarrow B^2\Sigma^+$ transition is near diagonal because the $3\sigma \rightarrow 4\sigma$ electron promotion from a non-bonding Si-centered orbital to a slightly anti-bonding Si-centered orbital results in an insignificant change in $R_e$. On the other hand, the $A^2\Pi_i$ (v) $\rightarrow B^2\Sigma^+$(v) transition is much less diagonal because the $R_e$ for the $A^2\Pi_i$ state is significantly longer than that for the $B^2\Sigma^+$ state since in the former the strongly bonding $1\pi$ orbital has an occupation of 3 whereas in the latter it is fully occupied. The predicted[29] permanent electric dipole moment of the $A^2\Pi_i$ state is much less than that for the $X^2\Sigma^+$ supporting the conjecture that the $3\sigma$ is Si-centered whereas the $1\pi$ is O-polarized bonding orbital.

Existing data can also be used to predict $b_{iv',fv''}$. The branching ratio, $b_{iv',fv''}$, is defined as[38],

$$b_{iv',fv''} = \frac{A_{iv',fv''}}{\sum_{fv''} A_{iv',fv''}} = \frac{\left(\left|\mu_{iv',fv''}\right|\right)^2 \times \left(\nu_{iv',fv''}\right)^3}{\sum_{fv''}\left(\left|\mu_{fv''}\right|\right)^2 \times \left(\nu_{iv',fv''}\right)^3} = \frac{I_{iv',fv''}}{\sum_{fv''} I_{iv',fv''}} \quad , \qquad 1)$$

where $A_{iv',fv''}$ is the Einstein coefficient for spontaneous emission, $\mu_{iv',fv''}$ $\equiv \left\langle \Psi_{v',J'} \left| \left\langle \Psi_{el'} \left| \hat{\mu}^{el}(r) \right| \Psi_{el'} \right\rangle \right| \Psi_{v'',J'} \right\rangle$ is the transition dipole moment, $\nu_{iv',fv''}$ is the transition frequency, and $I_{iv',fv''}$ is the intensity. The summation in Eq. 1 runs over all the lower vibronic states. If it is assumed that transition moment operator, $\hat{\mu}$ , is independent of internuclear separation (i.e. Condon approximation), then $b_{iv',fv''}$ is proportional to the product of the FCF and the $\nu^3$ factor. The radiative lifetime of an individual excited state vibronic level, $\tau_{iv'}$, is related to $A_{iv',fv''}$ simply by:

$$\tau_{iv'}^{-1} = \sum_{fv''} A_{iv',fv''} \qquad .$$

2)

Combination of these relationships gives[32]

$$\mu_{iv',fv''} = \sqrt{\frac{b_{iv',fv''}\tau_{iv'}^{-1}}{3.1362\times10^{-7} \nu_{iv',fv''}^3}} \quad ,$$

3)

where the conversion factor assumes that $\mu_{iv',fv''}$ is in Debye (D), $\nu_{iv',fv''}$ is in wavenumber (cm$^{-1}$) and $\tau_{iv'}$ is in seconds. Using the experimentally determined values for $b_{B\,^2\Sigma^+(v=0),fv''}$ based upon the aforementioned relative intensity measurements given in Table 1, and the experimentally determined $\tau_{iv'}$ (66±2 ns), then the $\left|\mu_{B\,^2\Sigma^+(v=0),fv''}\right|^{\text{exp.}}$ values given in Table 1 are obtained. The calculated transition dipole moments, $\left|\mu_{B\,^2\Sigma^+(v=0),fv''}\right|^{\text{calc.}}$, and FCFs are also given in Table 1. These values were obtained using the spectroscopic parameters[22] and predicted transition dipole moments[29]. Specifically, the suite of programs developed by Prof. Robert LeRoy (Waterloo University)[39] were used to calculate matrix elements of $\hat{\mu}^{\text{el}}(\text{r})$. The potential energy curves for the $X^2\Sigma^+$, $A^2\Pi_i$, and $B^2\Sigma^+$ states were predicted using the first-order Rydberg-Klein-Rees (RKR1) procedure with the aid of the previously determined molecular constants[22]. The potential energies curves were used as input to a program that numerically solved the radial Schrödinger equation (LEVEL 8.2 program). The same program calculated $\left\langle \Psi_{v',J'} \left| \left\langle \Psi_{el'} \left| \hat{\mu}^{\text{el}}(\text{r}) \right| \Psi_{el''} \right\rangle \right| \Psi_{v'',J''} \right\rangle$. A perusal of the *ab initio* prediction[29] reveals that electronic transition dipole moments, $\left\langle \Psi_{el'} \left| \hat{\mu}^{\text{el}}(\text{r}) \right| \Psi_{el''} \right\rangle$ are nearly linear in internuclear separation in the Franck-Condon region. For the purpose of prediction performed here, the slope and intercept for the $\left\langle B^2\Sigma^+ \left| \hat{\mu}^{\text{el}}(\text{r}) \right| X^2\Sigma^+ \right\rangle$ linear function were taken as -3.11D/Å and 6.57D, respectively, and for $\left\langle B^2\Sigma^+ \left| \hat{\mu}^{\text{el}}(\text{r}) \right| A^2\Pi_i \right\rangle$ as

-0.46D/Å and 0.84 D, respectively, which produced the $\left| \mu_{B^2\Sigma^+(v=0),f_{V''}} \right|^{\text{calc.}}$ values given in Table 1. The calculated branching ratios, $b_{iv',fv''}(\text{calc})$, were then obtained using Eq. 1. The obtained values are consistent with the previously predicted FCF[25] ratios which are also listed in Table 1.

The observed branching ratios, $b_{iv',fv''}(\text{exp.})$, indicate that the $B^2\Sigma^+(v=0) \rightarrow X^2\Sigma^+(v)$ emission is significantly less diagonal than that predicted (e.g. $\dfrac{b_{0,1}}{b_{0,0}}(\text{exp.}) = \dfrac{3.0}{97.0}$ vs. $\dfrac{b_{0,1}}{b_{0,0}}(\text{calc.}) = \dfrac{0.1}{99.6}$). This discrepancy may be due in part to the fact that the prediction does not account for the rotation and spin-orbit mixing of the $X^2\Sigma^+$ and $A^2\Pi_i$ states. Specifically, the spin-orbit mixing term, $\xi$, and the rotation mixing term, $2\eta$, have been determined[22] to be -37.5 cm$^{-1}$ and 0.85 cm$^{-1}$, respectively. Furthermore, the spectroscopic parameters used for the potential energy predictions were derived from rotational parameters of an effective Hamiltonian operator. Such parameters are contaminated with higher order terms used to account for Born Oppenheimer breakdown.

Finally, it is noteworthy that the *ab initio* predicted radiative lifetime[29], $\tau_{iv'}$ (= 59 ns), is near the experimentally determine values indicating that there is no significant non-radiative relaxation of the excited state levels. The $B^2\Sigma^+(v=0)$ rotational levels are embedded in a dense set of $A^2\Pi_i$ and $X^2\Sigma^+$ excited vibronic levels and could provide an avenue for non-radiative relaxation.

## V. Conclusion

We have demonstrated a method for producing a cold molecular beam of SiO$^+$ and performed the first dispersed laser induced fluorescence measurements to determine the radiative branching ratios of decay from the $B^2\Sigma^+(v=0)$ state. Decay to the $X^2\Sigma^+(v=0)$ state was observed to occur 97% of the time while the remaining 3% was observed to fall into $X^2\Sigma^+(v = 1)$. All other monitored decay channels were observed to have signals below detectability. Given the results of this study, we would expect an average of 30 spontaneous emissions on $B^2\Sigma^+$-$X^2\Sigma^+$ (0,0) before any off diagonal decay in $X^2\Sigma^+(v \neq 0)$ and even more before any decays into the $A^2\Pi_i$ manifold. At room temperature the rotational population density peaks near $N$=12. In regards to the previously mentioned rotational cooling technique, this corresponds to an average of 12 spontaneous emissions before population transfer to the lowest two rotational states. This suggests that efficient rotational cooling should be possible. A broadband vibrational re-pump will improve this efficiency, though the extent of the improvement depends significantly on branching rates to $A^2\Pi_i$ and, to a lesser extent, also depends on branching rates to higher vibrational states of $X^2\Sigma^+$.

Further investigation is necessary before decay paths and repump schemes can be fully characterized. For example, the Doppler cooling of $SiO^+$ proposed by Nguyen et al[18] assumes certain theoretical predictions of the $B^2\Sigma^+$ to $A^2\Pi_i$ branching fractions. A related possibility of performing single-molecule fluorescence detection of trapped $SiO^+$ also would require more scattered photons than does rotational cooling. Given a vibrational re-pump laser, the currently available data does not preclude this possibility, however the limits on the $B^2\Sigma^+$ to $A^2\Pi_i$ state branching fractions need to be improved by approximately two orders of magnitude to meaningfully assess the prospects of Doppler cooling. Similarly, design of a single molecule fluorescence apparatus would depend strongly on these branching fractions. Knowing the $A^2\Pi_i$ state radiative lifetime is also necessary to determine if the population will eventually become trapped in a dark state. We are currently working towards measuring the very small branching to the $A^2\Pi_i$ state to help answer such questions.

**Acknowledgements**

The research at ASU has been supported by a grant from the National Science Foundation (CSDM-A; CHE-1265885)(Steimle). NSF Grant No. PHY-1404455 funded PRS and BCO for hardware and development of software used for photon counting fluorescence detection. ARO Grant No. W911NF-14-0378 funded PRS and BCO for travel and data analysis. We thank Mark Kokish for useful conversations on molecular orbital theory.

**Reference**

1. L. D. Carr, D. DeMille, R. V. Krems and J. Ye, New J. Phys. **11** (May), No pp. given (2009).

2. G. Quemener and P. S. Julienne, Chem. Rev. (Washington, DC, U. S.) **112** (9), 4949-5011 (2012).

3. J. J. Hudson, D. M. Kara, I. J. Smallman, B. E. Sauer, M. R. Tarbutt and E. A. Hinds, Nature (London, U. K.) **473** (7348), 493-496 (2011).

4. J. F. Barry, D. J. McCarron, E. B. Norrgard, M. H. Steinecker and D. DeMille, Nature **512** (7514), 286-289 (2014).

5. D. J. McCarron, E. B. Norrgard, M. H. Steinecker and D. DeMille, New J. Phys. **17** (March), 1-12 (2015).

6. E. B. Norrgard, D. J. McCarron, M. H. Steinecker, D. DeMille and M. R. Tarbutt, Phys Rev Lett **116** (6), 063004 (2016).

7. M. T. Hummon, M. Yeo, B. K. Stuhl, A. L. Collopy, Y. Xia and J. Ye, Phys. Rev. Lett. **110** (14), 143001/143001-143001/143005 (2013).

8. V. Zhelyazkova, A. Cournol, T. E. Wall, A. Matsushima, J. J. Hudson, E. A. Hinds, M. R. Tarbutt and B. E. Sauer, Phys. Rev. A: At., Mol., Opt. Phys. **89** (5-B), 053416/053411-053416/053415 (2014).

9. C.-Y. Lien, C. M. Seck, Y.-W. Lin, J. H. V. Nguyen, D. A. Tabor and B. C. Odom, Nat. Commun. **5**, 4783 (2014).

10. M. Hamamda, P. Pillet, H. Lignier and D. Comparat, J. Phys. B: At., Mol. Opt. Phys. **48** (18), 1-24 (2015).

11. Y. Wan, F. Gebert, F. Wolf and P. O. Schmidt, Phys. Rev. A: At., Mol., Opt. Phys. **91** , 043425/043421-043425/043410 (2015).

12. P. O. Schmidt, T. Rosenband, C. Langer, W. M. Itano, J. C. Bergquist and D. J. Wineland, Science **309** (5735), 749-752 (2005).

13. F. Wolf, Y. Wan, J. C. Heip, F. Gebert, C. Shi and P. O. Schmidt, Nature **530** (7591), 457-460 (2016).

14. M. Viteau, A. Chotia, M. Allegrini, N. Bouloufa, O. Dulieu, D. Comparat and P. Pillet, Science **321** (5886), 232-234 (2008).

15. J. H. V. Nguyen, C. R. Viteri, E. G. Hohenstein, C. D. Sherrill, K. R. Brown and B. Odom, New J. Phys. **13** (June), 063023/063021-063023/063028 (2011).

16. C. Y. Lien, S. R. Williams and B. Odom, Phys Chem Chem Phys **13** (42), 18825-18829 (2011).

17. D. Sofikitis, S. Weber, A. Fioretti, R. Horchani, M. Allegrini, B. Chatel, D. Comparat and P. Pillet, New J Phys **11** (2009).

18. J. H. V. Nguyen and B. Odom, Phys Rev A **83** (5) (2011).

19. R. Cameron, T. J. Scholl, L. Zhang, R. A. Holt and S. D. Rosner, J. Mol. Spectrosc. **169** (2), 364-372 (1995).

20. T. J. School, R. Cameron, S. D. Rosner and R. A. Holt, Phys. Rev. A: At., Mol., Opt. Phys. **51** (3), 2014-2019 (1995).

21. T. J. Scholl, R. Cameron, S. D. Rosner and R. A. Holt, Can. J. Phys. **73** (1 & 2), 101-105 (1995).

22. S. D. Rosner, R. Cameron, T. J. Scholl and R. A. Holt, J. Mol. Spectrosc. **189** (1), 83-94 (1998).

23. R. Cameron, T. J. Scholl, L. Zhang, R. A. Holt and S. D. Rosner, J. Mol. Spectrosc. **169** (2), 352-363 (1995).

24. L. Zhang, R. Cameron, R. A. Holt, T. J. Scholl and S. D. Rosner, Astrophys. J. **418** (1 Pt. 1), 307-309 (1993).

25. R. R. Reddy, Y. Nazeer Ahammed, B. Sasikala Devi, K. Rama Gopal, P. Abdul Azeem and T. V. R. Rao, Astrophys. Space Sci. **281** (4), 729-741 (2002).

26. Z. L. Cai and J. P. Francois, Chem. Phys. **234** (1-3), 59-68 (1998).

27. Z. L. Cai and J. P. Francois, Chem. Phys. Lett. **282** (1), 29-38 (1998).

28. Z. L. Cai and J. P. Francois, J. Mol. Spectrosc. **197** (1), 12-18 (1999).

29. S. Chattopadhyaya, A. Chattopadhyay and K. K. Das, J. Mol. Struct.: THEOCHEM **639**, 177-185 (2003).

30. D. Shi, W. Li, W. Xing, J. Sun, Z. Zhu and Y. Liu, Comput. Theor. Chem. **980**, 73-84 (2012).

31. D. L. Kokkin, T. Ma, T. Steimle and T. J. Sears, J. Chem. Phys., Ahead of Print (2016).

32. D. L. Kokkin, T. C. Steimle and D. DeMille, Phys. Rev. A: At., Mol., Opt. Phys. **90** (6-A), 062503/062501-062503/062510 (2014).

33. N. J. Reilly, T. W. Schmidt and S. H. Kable, J. Phys. Chem. A **110** (45), 12355-12359 (2006).

34. J. R. Gascooke, U. N. Alexander and W. D. Lawrance, J. Chem. Phys. **134** (18), 184301/184301-184301/184314 (2011).

35. R.S. Ram , R. Engleman Jr., and P.F. Bernath, J. Mol. Spectrosc. **190**, 352-363 (1995).

36. Z. Cai and J. P. Francois, J Mol Spectrosc **197** (1), 12-18 (1999).

37.  S. Chattopadhyaya, A. Chattopadhyay and K. K. Das, J Mol Struc-Theochem **639**, 177-185 (2003).

38.  F. Arqueros and J. Campos, Physica B+C (Amsterdam) **112** (1), 131-137 (1982).

39.  R. J. LeRoy, University of Waterloo Chemical Physics Research Report CP-657R, 2004.

**Figure Caption**

**Figure 1.** (Color online) The 2D spectrum of the SiH₄/N₂O/Ar discharge covering 26000 cm⁻¹ to 26035 cm⁻¹ region of the $B^2\Sigma^+$ - $X^2\Sigma^+$ (0,0) transition. The vertical axis is the relative wavenumber of the dispersed fluorescence to the red of the wavenumber of excitation laser at the beginning of the scan. The viewed spectral region of the monochromator is tracked with laser wavelength. The features in the yellow rectangle are the on-resonance fluorescence of the $B^2\Sigma^+$- $X^2\Sigma^+$ (0,0) band of SiO⁺, the $^SR_{21f}(4.5)(\nu = 26008.3\ cm^{-1})$, $^SR_{21e}(4.5)(\nu = 26008.7\ cm^{-1})$, $^SR_{21f}(5.5)(\nu = 26026.5\ cm^{-1})$ and $^SR_{21e}(5.5)(\nu = 26026.9\ cm^{-1})$ lines of the $A^2\Delta$-$X^2\Pi_r$ (1,0) band of SiH and minor contributions from unknown species. The two intense, partially resolved, doublets 2050 cm⁻¹ red shifted from the laser are primarily due to the $^RR_{1f}(4.5)(\nu = 23963.1\ cm^{-1})$, $^RR_{1e}(4.5)(\nu = 23963.5\ cm^{-1})$, $^RR_{1f}(5.5)(\nu = 23969.5\ cm^{-1})$ and $^RR_{1e}(5.5)(\nu = 23970.3\ cm^{-1})$ transitions of the $A^2\Delta$ (v=1)→ $X^2\Pi_r$ (v=1) band.

**Figure 2**. Upper trace: The integration of the on-resonance horizontal slice indicated by the yellow rectangle of Figure 1. Lower trace: The simulated excitation spectrum of the $B^2\Sigma^+$- $X^2\Sigma^+$ (0,0) band of SiO⁺ obtained using a rotational temperature of 40 K and a linewidth of 0.4 cm⁻¹ FWHM. The line indicated by the arrow is due to an unidentified molecule that emits both on-resonance and 490 cm⁻¹ off-resonance (see Figure 1). The $P$(4), and higher branch members, of the $B^2\Sigma^+$- $X^2\Sigma^+$ (0,0) band are overlapped with emission from an with an unknown molecule.

**Figure 3**. (Color online) Dispersed laser-induced fluorescence (DLIF) spectra measured using (a) photon counting mode and (b) standard acquisition mode of ICCD. DLIF signal is the normalized average of 12 and 18 acquisitions of 15000 laser excitations for (a) and (b) respectively. Expected locations of $B^2\Sigma^+$(v=0) → $X^2\Sigma^+$(v″) and $B^2\Sigma^+$(v=0) → $A^2\Pi$ (v′) transitions are marked with dashed lines and show the only observable branching is weak branching to $X^2\Sigma^+$( v″=1). Variation of the baseline seen in the spectra taken with the standard acquisition can be explained by dark count variation from temperature drift in the CCD. The shape of the baseline is primarily a consequence of temperature gradients across the CCD. Intensifier gain allows the photon counting mode to be insensitive to dark count drifts, improving the SNR and eliminating the asymmetric baseline.

**Figure 4**. Radiative decay curves resulting from laser excitation of the $R$(3) line of the $B^2\Sigma^+$-$X^2\Sigma^+$(0,0) band and monitoring the on-resonance LIF signal. The dashed curve is the predicted decay obtained using the least-squares optimized lifetime.

Table 1. The branching ratio, $b_{iv',fv^*}$ and magnitude of the transition dipole moments, $\left|\mu_{iv',fv^*}\right|$, for the $B^2\Sigma^+$ ($N=3$, v'=0) $\rightarrow X^2\Sigma^+$ (v'') and $B^2\Sigma^+$ ($N=3$, v'=0) $\rightarrow A^2\Pi$(v'') transitions.

| | $X^2\Sigma^+$ | | | | $A^2\Pi$ | | |
|---|---|---|---|---|---|---|---|
| | v''=0 | v''=1 | v''=2 | v''=3 | v''=0 | v''=1 | v''=2 |
| $b_{iv',fv^*}$(exp) (%) | $97.0^{+0.7a}_{-2.5}$ | $3.0^{+0.7}_{-0.7}$ | <1.2[b] | <1.3[b] | <1.2[b] | <0.7[b] | <1[b] |
| $\left|\mu_{iv',fv^*}\right|^{exp.}$ (D) | $1.631^{+0.025c}_{-0.032}$ | $0.31^{+0.04}_{-0.04}$ | <0.21 | <0.23 | <0.21 | <0.17 | <0.22 |
| $\nu_{iv',fv^*}$ (cm⁻¹) | 26030 | 24870 | 23711 | 22552 | 23787 | 22847 | 21907 |
| $\left|\mu_{iv',fv^*}\right|^{calc.}$ (D) | 1.822[d] | 0.073 | 0.006 | 0.001 | 0.04 | 0.06 | 0.07 |
| FCFs$^{calc.}$ | 0.988 | 0.011 | 0.000 | 0.000 | 0.143 | 0.285 | 0.280 |
| FCFs[e] | 0.988 | 0.011 | 0.002 | 0.011 | 0.099 | 0.263 | 0.306 |
| $b_{iv',fv^*}$(calc) (%) | 99.6[f] | 0.1 | <0.1 | <0.1 | <0.1 | <0.1 | <0.1 |

[a] Estimated uncertainty comes from statistical uncertainty and limits on possible $B^2\Sigma^+ \rightarrow A^2\Pi$ relaxation (see text).

[b] Upper limit derived from noise floor at expected transition (see text).

[c] Obtained using Eq.1 and the measured values for using $b_{iv',fv^*}$ and $\tau$.

[d] Obtained using RKR potentials and the electronic *ab initio* transition moment form Ref. 29 (see text).

[e] Franck-Condon Factors from Ref. 23.

[f] From Eq. 1.

# Figure 1

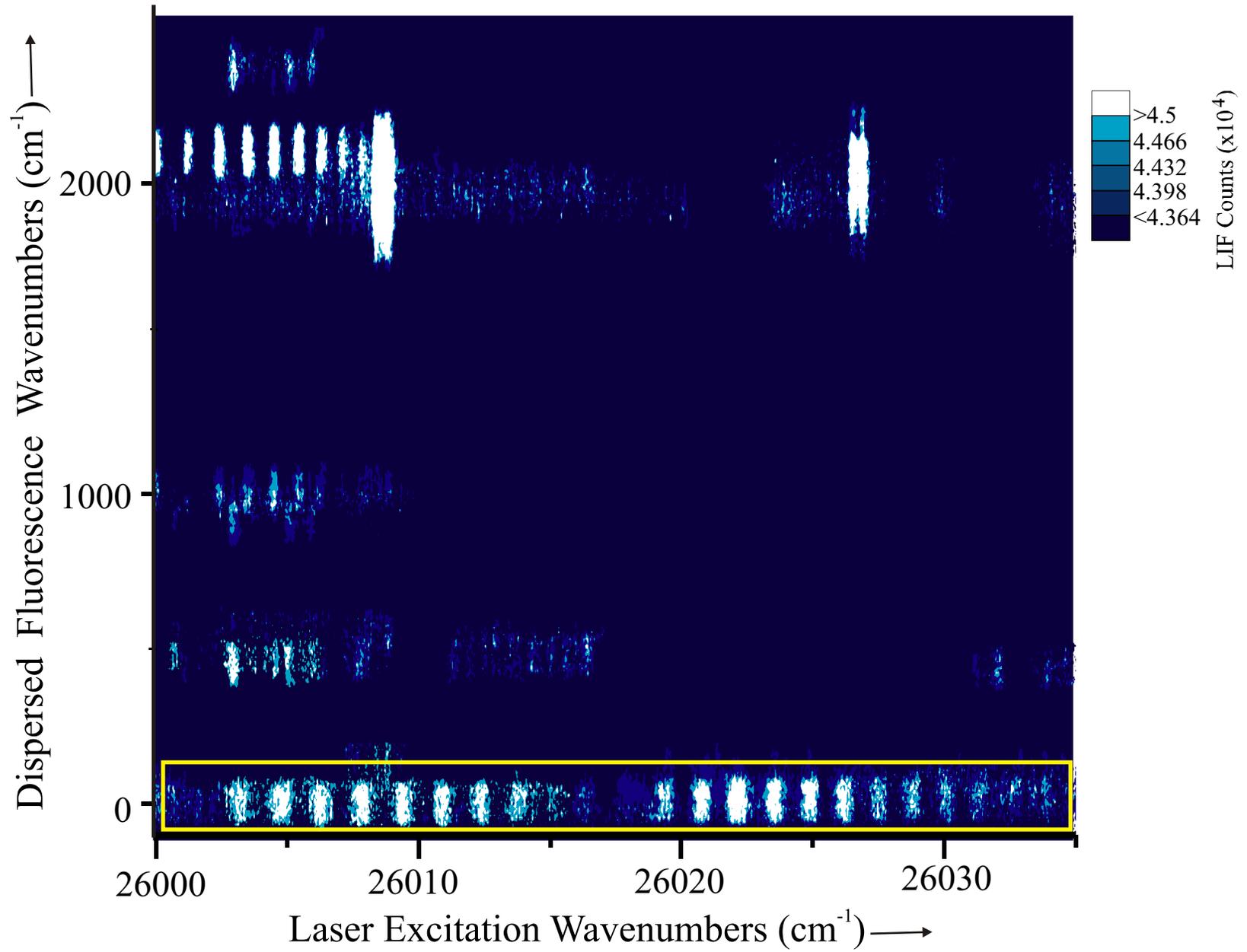

# Figure 2

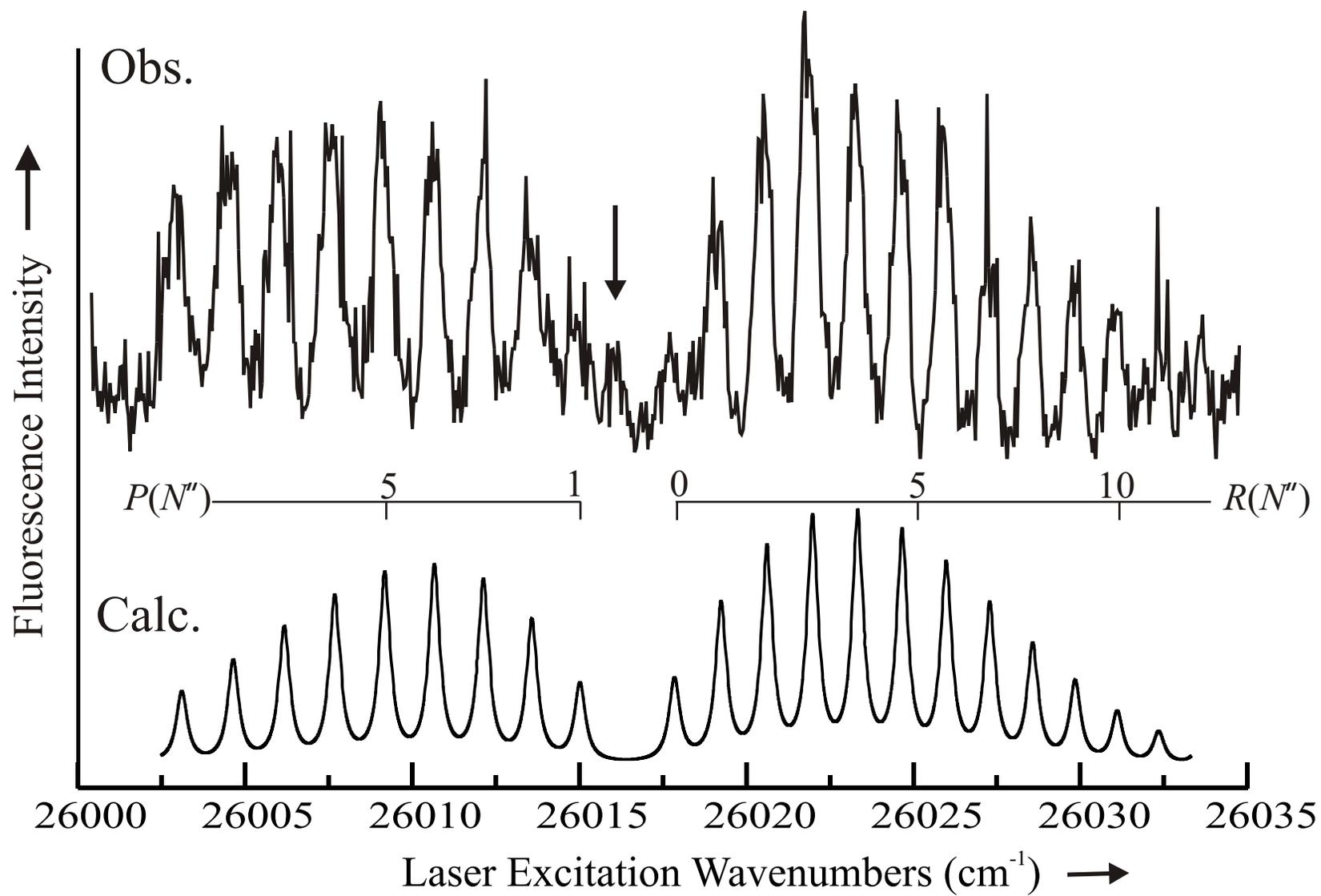

# Figure 3

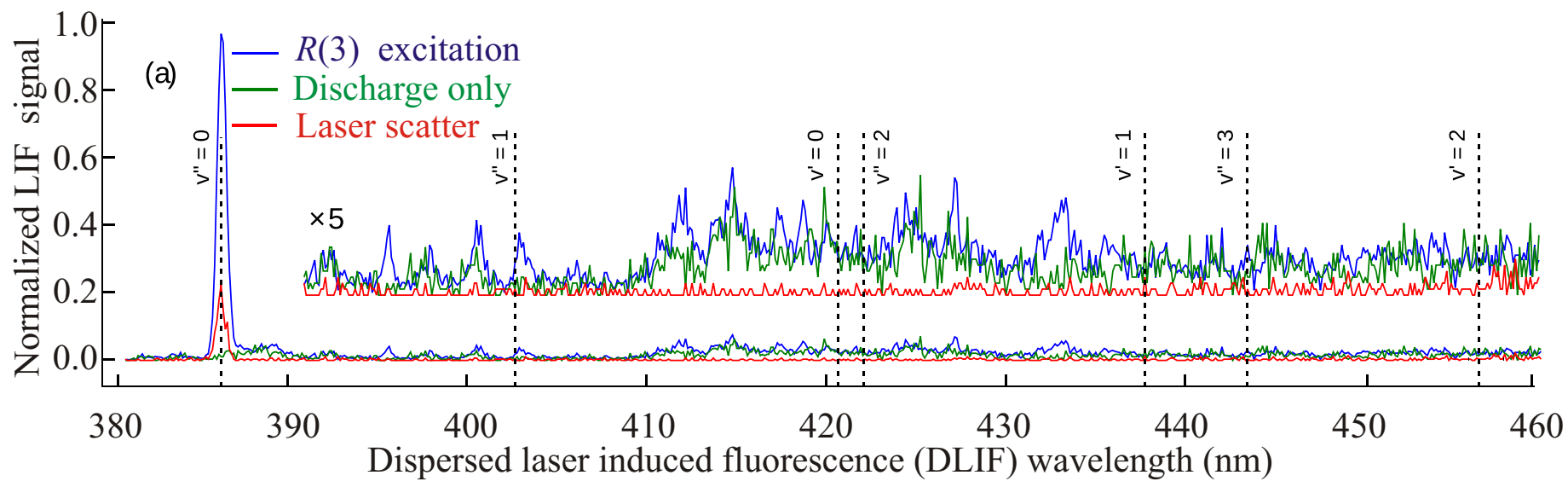

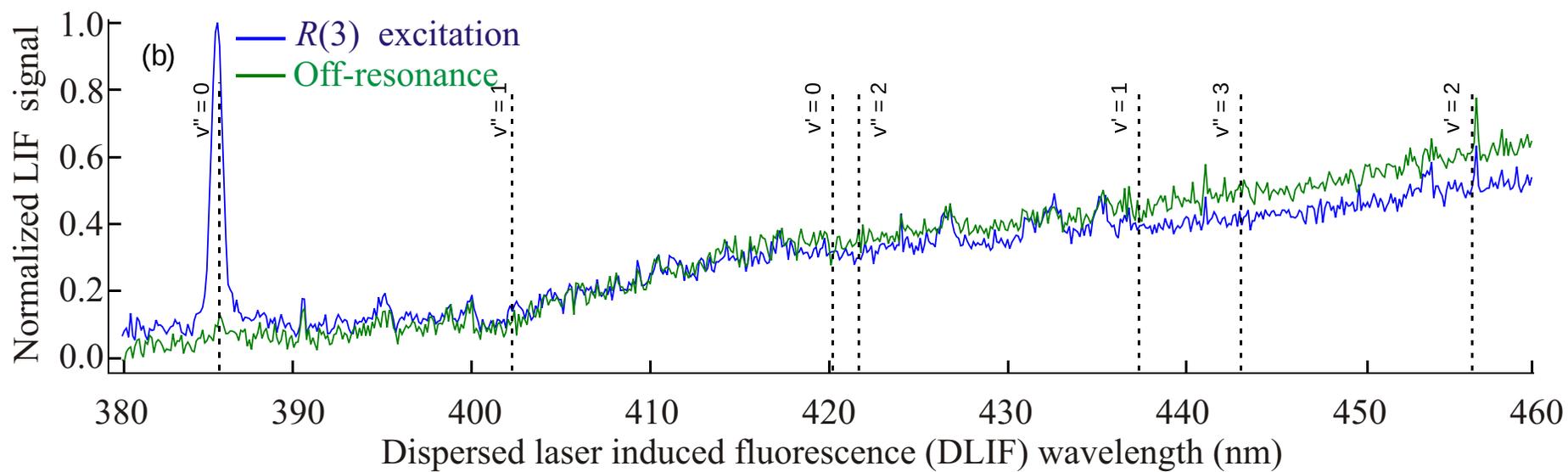

# Figure 4

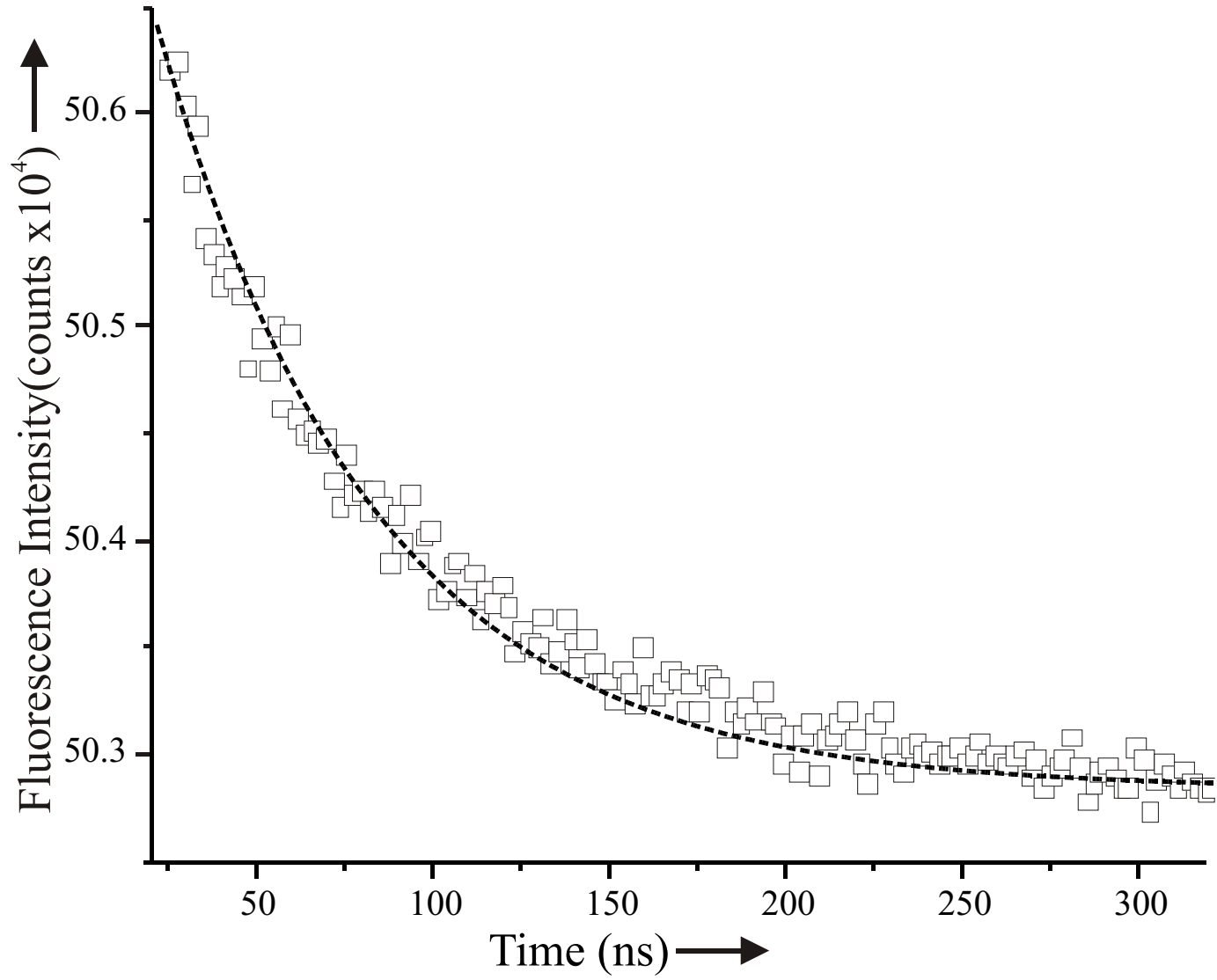